\documentclass[aps,pre,superscriptaddress,showpacs,amsmath,amssymb,twocolumn]{revtex4}

\usepackage[dvipdfmx]{graphicx}
\usepackage{bm}% bold math
\usepackage{color}
\usepackage{amsmath}
\usepackage{here}

\def\Vect#1{\mbox{\boldmath $#1$}}
\def\phij{\phi_{\rm c}}

\begin{document}

\title{Discontinuous change of shear modulus for frictional jammed granular materials
}

\author{Michio Otsuki}
\email[]{otsuki@riko.shimane-u.ac.jp}
\affiliation{Department of Physics and Materials Science, Shimane University, 1060 Nishikawatsu-cho, Matsue 690-8504, Japan}
\author{Hisao Hayakawa }
\affiliation{Yukawa Institute for Theoretical Physics, Kyoto University, Kitashirakawaoiwake-cho, Sakyo-ku, Kyoto 606-8502, Japan}

%\publishedin{% %Write this ONLY in cases of addenda and errata
%Prog.~Theor.~Phys.\ \textbf{XX} (19YY), page.}

%\recdate{Mmmmm DD, YYYY}% %Editorial Office will fill in this.

\begin{abstract}
The shear modulus of jammed frictional granular materials 
with the harmonic repulsive interaction
under an oscillatory shear
is numerically investigated.
It is confirmed that the storage modulus,
the real part of the shear modulus,
for frictional grains 
with sufficiently small strain amplitude $\gamma_0$
discontinuously emerges at the jamming transition point.
The storage modulus for small $\gamma_0$ differs from that 
of frictionless grains
even in the zero friction limit,
while they are almost identical with each other
for sufficiently large $\gamma_0$,
where the transition becomes continuous.
The stress-strain curve
exhibits a hysteresis loop even for a small strain,
which
connects a linear region for sufficiently 
small strain to another linear region
for larger strain.
We propose a new scaling law to interpolate between 
the states of small and large $\gamma_0$.
\end{abstract}
\date{\today}

\pacs{45.70.-n,05.70.Jk,81.40.Jj}

\maketitle

{\it Introduction.--}
When the packing fraction $\phi$ exceeds a critical value  $\phij$,
amorphous materials consisting of repulsive particles
such as granular materials, colloidal suspensions, foams, 
and emulsions turn into jammed solids
which have rigidity.
Such a transition, known as the jamming transition,
has been the subject of extensive studies 
over the last two decades \cite{Liu,Hecke}.
For frictionless grains,
the pressure increases continuously from $\phij$,
while the coordination number $Z$
exhibits a discontinuous transition at $\phij$ in the hard core
limit \cite{OHern02,OHern03}.

An assembly of frictionless grains under 
a simple shear
exhibits a rheological continuous transition:
the viscosity diverges
as $\phi$ approaches $\phij$ below,
while the yield stress increases continuously above $\phij$
\cite{Olsson, Hatano07, Hatano08, Tighe, Hatano10, Otsuki08, Otsuki09, Otsuki10, Nordstrom, Olsson11, Vagberg, Otsuki12, Ikeda, Olsson12, DeGiuli, Vagberg16,
Boyer, Trulsson, Andreotti, Lerner, Vagberg14, Kawasaki15, Suzuki}.
The jamming transition is also characterized by
the appearance of rigidity
under an oscillatory strain above $\phij$.
For sufficiently small strain,
the elastic modulus of frictionless grains is independent of the strain
and
 the critical exponents 
 for the jamming transition depend on the type of the local interaction
\cite{OHern02,OHern03,Wyart05}.
We call this regime the linear response regime.
For large strain,
recent studies
\cite{Coulais,Otsuki14,Deen,Goodrich,Nakayama,Boschan}
have revealed that 
the real part of the shear modulus,
the storage modulus $G$, of frictionless particles decreases with increasing strain 
as a result of nonlinear response
because of slip avalanches
\cite{Dahmen98,Dahmen11}.
The present authors have proposed a scaling law of $G$
to interpolate between the linear and the nonlinear responses \cite{Otsuki14}.
Note that the storage modulus 
as the ratio to the stress to the 
strain can be used even in the nonlinear response regime.

Most of previous studies, however, assume
that grains are frictionless,
though it is impossible to remove contact friction
in experiments of dry granular particles and
the friction causes drastic changes in rheology
such as a discontinuous shear-thickening with
a hysteresis loop near $\phij$
\cite{Otsuki11,bob_nature,Chialvo,Brown,Seto,Fernandez,Heussinger,Bandi,Ciamarra,Mari,Grob,Kawasaki14,Wyart14,Grob16,Hayakawa,Magnanimo}.
Somfai et al. \cite{Somfai} have numerically investigated 
the elastic moduli of a frictional system with 
the aid of
the density of state,
and found that
$G/B$ with the bulk modulus $B$
in the linear response regime
 is proportional to $\Delta Z$,
the excess coordination number relative to the isostatic value.
They have also found that $\Delta Z$
discontinuously appears
at $\phij$ 
for frictional grains \cite{Somfai}.
This suggests that
frictional grains with the harmonic repulsive interaction exhibit
a discontinuous change of $G$ at $\phij$
even in the frictionless limit
contrast to a continuous change 
for frictionless grains \cite{OHern02,OHern03,Wyart05}.

The difference of the linear elasticity
between the frictionless limit and the frictionless case
casts doubt on the expectation that
the scaling laws for the elasticity of frictionless grains 
can be confirmed in experiments
of grains with sufficiently small friction coefficient.
The accessible shear strain 
in the experiment \cite{Coulais}, however, is too large, 
and does not correspond to 
the previous theoretical studies \cite{OHern02,OHern03,Otsuki14}.
Also, little is known on
the nonlinear elasticity of jammed frictional grains.
To clarify the effects of the contact friction
on the linear and the nonlinear elasticities,
we numerically investigate the shear modulus
of two-dimensional frictional granular materials 
near the jamming point
under an oscillatory shear.

%Setup
{\it Setup of Simulation.--}
Let us consider a two-dimensional assembly of $N$ frictional granular particles.
They interact according to
the Cundall-Strack model
with an identical mass density $\rho$
in a square periodic box of linear size $L$
\cite{Cundall}.
The normal repulsive interaction force $F^{\rm (n)}$
is given by $F^{\rm (n)}=k^{\rm (n)} r$
with the compression length $r$
and the normal spring constant $k^{\rm (n)}$,
while the tangential contact force $F^{\rm (t)}$ in quasi-static motion
is constrained by the Coulomb criterion
$|F^{\rm (t)}| \le \mu F^{\rm (n)}$:
$F^{\rm (t)} = k^{\rm (t)} \delta^{\rm (t)}$
in the ``stick region'' for $|\delta^{\rm (t)}| < \mu F^{\rm (n)}/k^{\rm (t)} $
with the tangential spring constant $k^{\rm (t)}$
and
the tangential displacement $\delta^{\rm (t)}$,
while $|F^{\rm (t)}|$ remains $\mu F^{\rm (n)}$
in the ``slip region'' for $|\delta^{\rm (t)}| \ge \mu F^{\rm (n)}/k^{\rm (t)}$.
To avoid crystallization, we use a bi-disperse
system which includes
equal number of grains of the diameters $d_0$
and $d_0/1.4$, respectively.
The system is subjected to an oscillatory shear with
the shear strain $
\gamma (t) = \gamma_0 \left \{ 1 - \cos(\omega t) \right \}$,
where $\gamma_0$ and $\omega$ are the strain amplitude and the angular frequency, respectively.
Then, we measure 
the storage modulus defined by \cite{Doi}
\begin{eqnarray}
G(\gamma_0,\mu,\phi) = - \frac{\omega}{\pi} \int_0^{2\pi/\omega} dt
\sigma(t) \cos(\omega t)/\gamma_0,
\end{eqnarray}
where $\sigma(t)$ is the shear stress.
Note that the loss modulus exhibits only a linear 
dependence on $\omega$ and its $\phi$-dependence is 
relatively small \cite{Supple}.
We also point out that $G$
is almost independent of $\omega$
when the time period of the oscillatory shear 
is sufficiently larger than the relaxation time
of the configuration of the grains.
We, thus, focus on the dependence of $G$
only on $\gamma_0$, $\mu$, and $\phi$ for sufficiently small $\omega$.
Further details of our model and
the $\omega$-dependence of the storage and the loss moduli
are shown in Ref. \cite{Supple}.

\begin{figure}[htbp]
\includegraphics[width=\linewidth]{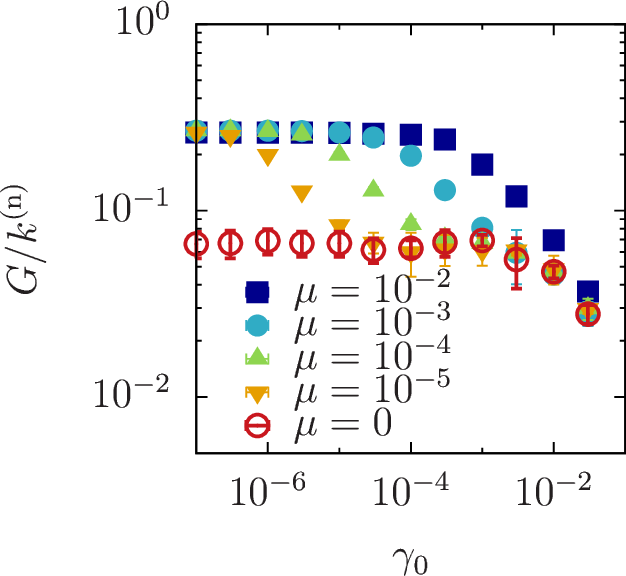}
\caption{(Color online) The storage modulus $G$ against $\gamma_0$ 
for $\mu=10^{-2}$, $10^{-3}$, $10^{-4}$, $10^{-5}$, and $0$ 
at $\phi=0.87$.
  }
\label{G_ga_phi0.87}
\end{figure}

{\it Storage modulus for a given packing fraction.--}
To begin with,
we study the dependence of the storage modulus
on the friction coefficient $\mu$.
In Fig. \ref{G_ga_phi0.87},
we plot $G$ against $\gamma_0$ 
for various $\mu$ 
at $\phi=0.87$.
For each $\mu$,
$G$ is almost independent of $\gamma_0$ 
in the linear response regime,
while $G$ decreases with $\gamma_0$
  in the nonlinear response regime.
In the linear response regime, $G$ for $\mu>0$
is almost independent of $\mu$, but differs from that for $\mu=0$ \cite{Somfai}.
In the nonlinear response regime,
the storage moduli for $\mu=0$ and $\mu>0$ are almost identical with each other.
It is noteworthy that the range of the linear response 
becomes narrower as $\mu$ decreases, and $G$ for $\mu=10^{-5}$ and $10^{-4}$
have second plateaus.
We confirm that storage moduli for various $\phi$
depend on the order of the limits:
\begin{equation}
\lim_{\mu \to +0} \lim_{\gamma_0 \to +0} G(\gamma_0,\mu,\phi)
\neq \lim_{\gamma_0 \to +0} G(\gamma_0,\mu=0,\phi).
\label{limit:1}
\end{equation}

\begin{figure}[htbp]
\includegraphics[width=\linewidth]{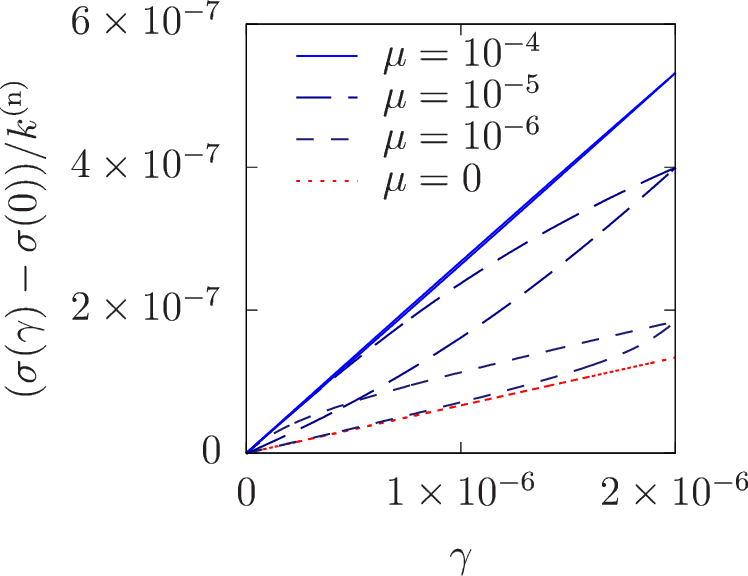}
\caption{
(Color online) The intrinsic shear stress $\sigma(\gamma) - \sigma(0)$ against $\gamma$ 
with $\mu =10^{-4}$, $10^{-5}$, $10^{-6}$, and $0$
for $\gamma_0=10^{-6}$ and $\phi = 0.87$.
}
\label{s_ga1.0e-06}
\end{figure}

The dependence of $G$ on $\mu$ and $\gamma_0$ in Fig. \ref{G_ga_phi0.87}
 can be explained from 
the stress-strain curves shown in Fig. \ref{s_ga1.0e-06},
where we plot 
the intrinsic shear stress $ \sigma(\gamma) - \sigma(0)$ against 
the shear strain $\gamma$ 
for various $\mu$
at $\gamma_0=10^{-6}$ and $\phi=0.87$.
For each $\mu$, 
$\sigma(\gamma) - \sigma(0)$ is proportional to $\gamma$
for sufficiently small $\gamma$.
The proportionality constant for $\mu>0$
is independent of $\mu$, but is larger than that for $\mu=0$,
which means Eq. \eqref{limit:1}.
The range of the linear response becomes narrower as $\mu$ decreases, and 
the stress-strain curve at $\mu=10^{-6}$ exhibits
a hysteresis loop which connects
 the first linear region for
sufficiently small $\gamma$ with the second linear region.
The gradient for the second linear region is identical to that
of the frictionless grains,
which results in the second plateau shown in Fig. \ref{G_ga_phi0.87}.
This behavior has its origin in the change
of the tangential friction $F^{\rm (t)}$
from the ``stick region'' to the ``slip region''.
See Ref. \cite{Supple} for 
the $\gamma_0$-dependence of the stress-strain curve.

\begin{figure}[htbp]
\includegraphics[width=\linewidth]{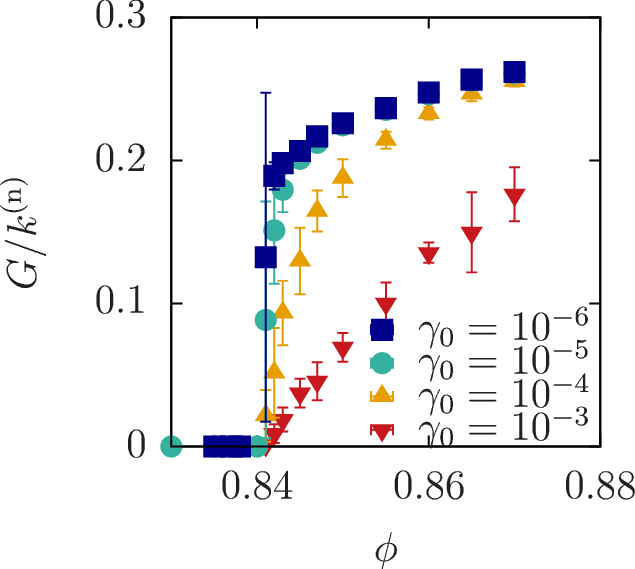}
\caption{(Color online) The storage modulus $G$ against $\phi$ 
for $\gamma_0=10^{-6}$, $10^{-5}$, $10^{-4}$, and $10^{-3}$ at $\mu=0.01$.
}
\label{G_phi}
\end{figure}

{\it Scaling law.--}
Next, we examine how the storage modulus depends on $\phi$.
In Fig. \ref{G_phi}, we plot 
$G$ against $\phi$ for various $\gamma_0$ 
at $\mu=0.01$.
For the smallest strain amplitude ($\gamma_0=10^{-6}$), 
$G$ exhibits a discontinuous
transition at $\phij \simeq 0.84$. 
As $\gamma_0$ increases,
the discontinuity at $\phij$ decreases and
the transition
becomes asymptotically continuous,
where $G$ is approximately proportional to $\phi - \phij$.

\begin{figure}[htbp]
\includegraphics[width=\linewidth]{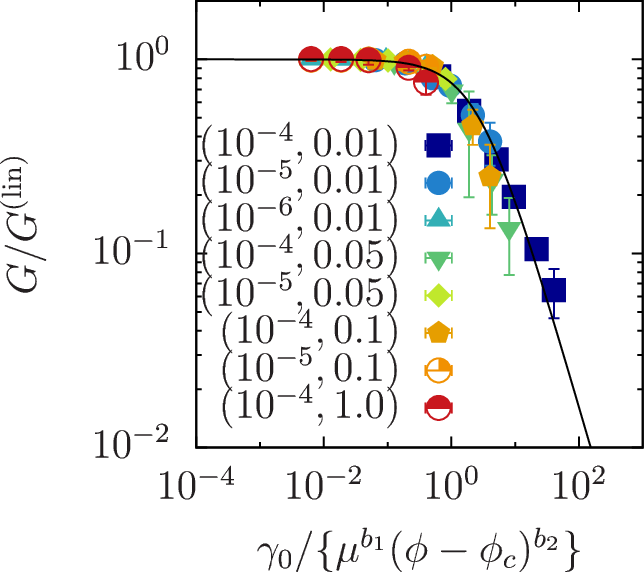}
\caption{(Color online) Scaling plot of $G$ characterized by Eq. \eqref{scale:eq}.
Each symbol in the legend is characterized by
$(\gamma_0,\mu)$,
but each one has the data
for $\phi-\phij=0.0001, 0.0002, 0.0005, 0.001$, and $0.01$.
The solid line represents the scaling function (S16) in Ref. \cite{Supple}.}
\label{G_p_scale}
\end{figure}

Here, we propose a new scaling law for the storage modulus
near the transition point:
\begin{equation}
G(\gamma_0,\mu,\phi) = G^{\rm (lin)}(\mu,\phi)
{\mathcal F}\left ( \frac{\gamma_0}{\mu^{b_1} \{\phi - \phij(\mu)\}^{b_2}}\right ),
\label{scale:eq}
\end{equation}
where ${\mathcal F}(x)$ is the scaling function satisfying
\begin{equation}
\lim_{x\to0}{\mathcal F}(x) = 1, \ \ \lim_{x\to\infty}{\mathcal F}(x) \sim x^{-c}
\label{F}
\end{equation}
with exponents $b_1$, $b_2$, and $c$,
and we have introduced 
\begin{equation}
G^{\rm (lin)}(\mu,\phi) \equiv \lim_{\gamma_0 \to +0} G(\gamma_0,\mu,\phi).
\label{Glin:eq}
\end{equation}
In Eq. \eqref{scale:eq}, we have used the jamming point $\phij(\mu)$ 
depending on $\mu$ \cite{Supple}.
Figure \ref{G_p_scale}
confirms the validity of the scaling plot characterized by Eq. \eqref{scale:eq},
where we have determined
$G^{\rm (lin)}(\mu,\phi)$ by 
the extrapolation of the limit $\gamma_0\to+0$
using the data for $\gamma_0 \ge 10^{-7}$.
The critical exponents used in Eq. \eqref{scale:eq},
are given by \cite{Supple}
\begin{equation}
b_1 = 1.00 \pm 0.02, \ \
b_2 = 0.90 \pm 0.02, \ \ c = 1.13 \pm 0.02.
\label{b}
\end{equation}
It should be noted that the scaling law
cannot be applied to the region of the second plateau in Fig. \ref{G_ga_phi0.87}

In Eq. \eqref{scale:eq},
we have assume that the critical strain $\gamma_c$
characterizing the crossover
from the linear 
to the nonlinear response regimes
is proportional to $\mu^{b_1} (\phi - \phij)^{b_2}$.
The exponents $b_1$ and $b_2$ in Eq. \eqref{b}
may be understood as follows.
First, $\gamma_c$ is expected to satisfy
$\gamma_c \sim \delta^{\rm (t)}_c$
with the critical tangential displacement $\delta^{\rm (t)}_c$
characterizing the change of the tangential friction $F^{\rm (t)}$
to the ``slip region''.
Then, we deduce $\delta^{\rm (t)}_c \sim \mu F^{\rm (n)}$
with the average contact force 
$F^{\rm (n)} \sim (\phi - \phij)$
for grains with the harmonic repulsive interaction
\cite{Otsuki08,Otsuki09}.
Thanks to the above relations,
we obtain
$\gamma_c \sim \mu^{b_1} (\phi - \phij)^{b_2}$
with $b_1 = 1$ and $b_2 = 1$, 
which are not far away from the estimated values in Eq. \eqref{b}.
We, however, do not identify
the reason why 
the evaluation of $b_2$ deviates a little from the numerical value. 

It should be noted that
the scaling form of Eq. \eqref{scale:eq}
is analogous to that for frictionless case
proposed in Ref. \cite{Otsuki14}, though
the $\mu$-dependence is not included and $c$ is 
$1/2$ in the conventional scaling.
The main difference between Eq. \eqref{scale:eq}
and the conventional one is that Eq.  \eqref{scale:eq}
represents the crossover 
from the stick to the slip branch,
 while the previous scaling deals with the crossover 
   from the slip 
to the avalanche branch.

\begin{figure}[htbp]
\includegraphics[width=\linewidth]{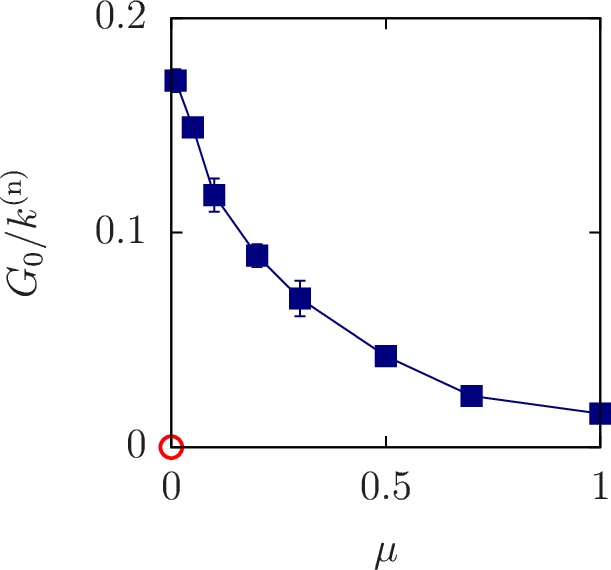}
\caption{(Color online) The storage modulus $G_0(\mu)$ for $\mu>0$
at $\phij$ in the linear response region.
The smallest value of $\mu$ in this plot
is $0.01$.
The open circle at the origin represents the result for frictionless grains.
}
\label{G0}
\end{figure}

We have also confirmed that 
the storage modulus $G^{\rm (lin)}(\mu,\phi)$ 
in the linear response regime exhibits
a discontinuous transition at $\phij$.
To give further evidence, we plot the storage modulus 
at $\phij$ defined as
\begin{equation}
G_0(\mu) \equiv \lim_{\phi \to +\phij} 
G^{\rm (lin)}(\mu,\phi)
\end{equation}
against $\mu$ in Fig. \ref{G0},
where $G_0(\mu)$ has a maximum value in the frictionless limit.
We have also found that
$G^{\rm (lin)}(\mu,\phi)$ satisfies
\begin{equation}
G^{\rm (lin)}(\mu,\phi) - G_0(\mu)
\propto \{ \phi - \phij(\mu) \}^{a}
\label{scaling2}
\end{equation}
with an exponent $a =  0.52 \pm 0.01$.
It should be noted that
the previous numerical results suggest $a = 1/2$
 \cite{Somfai,Supple}.

We have summarized our results by
two scaling laws derived from Eqs. \eqref{scale:eq} and \eqref{scaling2}:
\begin{equation}
\lim_{\gamma_0 \to +0} G(\gamma_0, \mu, \phi)
= G_0(\mu) + A(\mu) \{ \phi - \phij(\mu) \}^{a}
\label{scaling:3}
\end{equation}
for the discontinuous transition
in the linear response regime
and 
\begin{equation}
\lim_{\phi \to +\phij} G(\gamma_0, \mu, \phi)
\propto G_0(\mu) 
\left (\frac{\mu^{b_1} (\phi - \phij(\mu))^{b_2}}{\gamma_0} \right )^{c}
\label{scaling:4}
\end{equation}
for the continuous transition in the nonlinear response regime,
where $A(\mu)$ depends only on $\mu$.
We note that the scaling law, Eq. \eqref{scaling:4},
with $b_2=0.90$ and $c\simeq1.13$ indicates
$G \sim (\phi-\phij)^{b_2 c} \simeq (\phi-\phij)$
in the nonlinear response regime.
This might suggest that
the scaling of $G$ is identical to that of the pressure $P$
\cite{Otsuki14}.

% omega-dependence and loss modulus
{\it Discussion.--}
Let us discuss our results.
Tighe reported that the complex shear modulus $G^*$ 
for a model of emulsions
 satisfies
$G^* \sim \omega^{1/2}$ near $\phij$
in the linear response regime  \cite{Tighe11}.
On the contrary, 
we have confirmed that
the storage modulus is independent of $\omega$,
and the loss modulus is proportional to $\omega$
in the linear response regime
for all of the packing fractions \cite{Supple}.
It should be noted that 
the static contact friction and the inertia of the particles
exist in our system, 
but they are not involved in Ref. \cite{Tighe11},
which might lead to the different dependence of $G^*$.

% different contact force
In this Letter, we have proposed the scaling law for the shear modulus
of grains with the harmonic repulsive interaction.
From the analogy of the frictionless case \cite{Otsuki14},
we expect the exponent $b_2 = 3/2$ 
for Heltzian contact model
because of the relation $\gamma_c \sim \delta^{\rm (t)}_c \sim \mu F^{\rm (n)}$
with $F^{\rm (n)} \sim (\phi - \phij)^{3/2}$.
The confirmation of this conjecture
will be the subject of further study.

% Dimensional dependence and percolation

There are some studies to focus on
the role of percolation to the jamming 
\cite{Lois, Shen, Kovalcinova, Henkes}.
In the system exhibiting a conventional percolation,
the critical exponents for the storage modulus
depends on the spatial dimension  \cite{Torquato}, but the exponents of the jamming transition
are independent of the dimensionality, at least, for frictionless grains \cite{Otsuki09}. 
Further careful study on the relation between
percolation and 
jamming should be necessary.

An important finding of this Letter
is the scaling law \eqref{scale:eq}
interpolating between the linear and the nonlinear
response regimes
for frictional grains.
Somfai et al. found that
the elastic moduli
of frictional grains
in the linear response regime
is proportional to the excess coordination number $\Delta Z$ \cite{Somfai}.
Based on this result,
a recent review paper suggested that
there is no scaling of the mechanical properties 
as a function of $\phi - \phij$
for frictional grains
because $\Delta Z$ of the frictional grains
remains finite even at $\phij$ \cite{Hecke}.
Our scaling plot based on Eq. \eqref{scale:eq},
however, 
gives a counter evidence of the existence
of the scaling law.
We also note that Ref. \cite{Somfai} have ignored the change
of the tangential contact force to the ``slip region'',
but this effect becomes significant
in the vicinity of $\phij$ for finite $\gamma_0$
as can be seen in Eq. \eqref{scale:eq}.
To our knowledge,
this effect has not been indicated in previous studies.

{\it Concluding remarks.--}
We have numerically investigated the frictional granular particles.
The storage modulus 
in the linear response regime for frictional grains 
differs from that for frictionless grains even in the zero friction limit,
whereas they are almost identical in the nonlinear response regime.
This dependence on the tangential friction 
has been explained from the stress-strain curve 
shown in Fig. \ref{s_ga1.0e-06}.
We have also proposed the new scaling law
that interpolates between the discontinuous transition
for infinitesimal strain and the continuous one
for finite strain.
This scaling law has been verified through our simulation.

%\begin{acknowledgments}
The authors thank O. Dauchot, H. Yoshino, T. Yamaguchi, 
F. van Wijland,
and K. Miyazaki
 for fruitful discussions. 
This work is partially supported by the Grant-in-Aid of MEXT for Scientific 
Research (Grant No. 16H04025 and No. 25800220).
One of the authors (M.O.) appreciates the warm hospitality of Yukawa Institute for Theoretical Physics at Kyoto University and the discussions during the YITP workshops
 ``New Frontiers in Non-equilibrium Physics 2015'' and ``Avalanches, plasticity, and nonlinear response in nonequilibrium'', which helped to complete this work.
%\end{acknowledgments}

\setcounter{equation}{0}
\setcounter{figure}{0}

\renewcommand{\theequation}{S\arabic{equation}}
\renewcommand{\thefigure}{S\arabic{figure}}
\renewcommand{\bibnumfmt}[1]{[S#1]}
\renewcommand{\citenumfont}[1]{S#1}

\newpage

\begin{center}
\textbf{\Large Supplemental Materials: **}
\end{center}

\section{Introduction}
In this Supplemental Materials,
we present details
and additional results of our model.
We explain the setup of our simulation
in Sec. \ref{Details}.
In Sec. \ref{Omega},
we demonstrate the dependence of the complex
shear modulus on the angular frequency $\omega$.
Section \ref{Stress} deals with 
the stress-strain curves for various $\gamma_0$.
In Sec. \ref{Exp}, we explain the method
to determine the exponents for the scaling law
proposed in the main text.
We present the method to determine
the transition point depending on 
the friction coefficient $\mu$
in Sec. \ref{Phi}.
In Sec. \ref{Linear},
we show numerical results for the storage modulus in the linear response
regime.

\section{Details of our simulation}
\label{Details}

In this section, we explain our setup and model.
The position, the velocity, and the angular velocity
of the grain $i$ are, respectively, denoted by
$\Vect{r}_i$, $\Vect{v}_i$, and $\omega_i \hat{\Vect{z}}$,
where we have introduced the unit vector $\hat{\Vect{z}}$
parallel to $z$ axis
(perpendicular to the considering plane).
The contact force $\Vect{F}_{ij}$ 
between the grain $i$ and the grain $j$ 
consists of
the normal part $\Vect{F}^{\rm (n)}_{ij}$ 
and the tangential part $\Vect{F}^{\rm (t)}_{ij}$
as $\Vect{F}_{ij} = \Vect{F}^{\rm (n)}_{ij} + \Vect{F}^{\rm (t)}_{ij}$.
The normal contact force $\Vect{F}^{\rm (n)}_{ij}$ 
is given by
\begin{equation}
\Vect{F}^{\rm (n)}_{ij} = 
\Vect{F}^{\rm (n, el)}_{ij} +\Vect{F}^{\rm (n, vis)}_{ij},
\end{equation}
where 
\begin{eqnarray}
\Vect{F}^{\rm (n, el)}_{ij} & = & k^{\rm (n)} (d_{ij} - |\Vect{r}_{ij}|) 
\Theta (d_{ij} - |\Vect{r}_{ij}|) \Vect{n}_{ij}, \label{Flin}\\
\Vect{F}^{\rm (n, vis)}_{ij} & = & - \eta^{(n)} v^{\rm (n)}_{ij} \Theta (d_{ij} - |\Vect{r}_{ij}|) \Vect{n}_{ij}
\label{Flint}
\end{eqnarray}
with the normal spring constant $k^{(n)}$ and
the normal viscous constant $\eta^{(n)}$.
Here, $\Vect{r}_{ij}$, $\Vect{n}_{ij}$, $v^{\rm (n)}_{ij}$, and $d_{ij}$ 
are, respectively, given by
$\Vect{r}_{ij} = \Vect{r}_i - \Vect{r}_j$, 
$\Vect{n}_{ij} = \Vect{r}_{ij}/|\Vect{r}_{ij}|$,
$v^{\rm (n)}_{ij} = (\Vect{v}_{i}- \Vect{v}_{j}) \cdot \Vect{n}_{ij}$,
and
$d_{ij} = (d_i + d_j)/2$
with the diameter $d_i$ of the grain $i$.
In Eqs. \eqref{Flin} and \eqref{Flint},
we have introduced the
the Heaviside step function $\Theta(x)$ defined by $\Theta(x)=1$
for $x \ge 0$ and $\Theta(x)=0$ otherwise.

Similarly, the tangential friction force
$\Vect{F}^{\rm (t)}_{ij}$ is given by
\begin{equation}
\Vect{F}^{\rm (t)}_{ij} = \min \left ( \tilde F^{\rm (t)}_{ij}, \mu 
|\Vect{F}^{\rm (n, el)}_{ij}| \right ) 
\mathrm{sign} \left (\tilde F^{\rm (t)}_{ij} \right ) \Vect{t}_{ij},
\label{Coulomb}
\end{equation}
where $\min(a,b)$ selects the smaller one between $a$ and $b$,
 $\mathrm{sign}(x)$ is $1$ for $x>0$ and 
  $-1$ for $x<0$,
and $\tilde F^{\rm (t)}_{ij}$ is given by
\begin{equation}
\tilde F^{\rm (t)}_{ij} = k^{\rm (t)} \delta^{\rm (t)}_{ij} - \eta^{\rm (t)} v^{\rm (t)}_{ij}
\label{Ft:eq}
\end{equation}
with $\Vect{t}_{ij} = (-r_{ij,y}/|\Vect{r}_{ij}|, r_{ij,x}/|\Vect{r}_{ij}|)$.
Here, $k^{\rm (t)}$, $\eta^{\rm (t)}$, and $\mu$ are the tangential spring constant, the tangential viscous constant, 
and the friction coefficient, respectively.
The tangential velocity $v^{\rm (t)}_{ij}$ and the tangential displacement
$\delta^{\rm (t)}_{ij}$ are, respectively, given by
\begin{eqnarray}
v^{\rm (t)}_{ij} & = & (\Vect{v}_{i}- \Vect{v}_{j}) \cdot \Vect{t}_{ij}
- (d_i \omega_i + d_j \omega_j)/2, \\
\delta^{\rm (t)}_{ij} & = & \int_{\mathrm{stick}} dt \ v^{\rm (t)}_{ij},
\label{vt}
\end{eqnarray}
where ``stick'' on the integral implies that
the integral is only performed 
when the condition $|\tilde F^{\rm (t)}_{ij}| < \mu |\Vect{F}^{\rm (n, el)}_{ij}|$
is satisfied.
Additionally, we have introduced the torque $T_i$ on the grain $i$ as
\begin{equation}
T_i = - \sum_j \frac{d_i}{2} \Vect{F}^{\rm (t)}_{ij} \cdot \Vect{t}_{ij}.
\end{equation}

%In Fig. \ref{Ft}, we plot the tangential contact force $|\Vect{F}^{\rm (t)}_{ij}|$
% normalized with the normal contact force $|\Vect{F}^{\rm (n)}_{ij}|$
%against the tangential displacement $\delta^{\rm (t)}_{ij}$ normalized with
%$|\Vect{F}^{\rm (n)}_{ij} |/ k^{\rm (t)}$ for $\mu = 0.1$, $0.5$, $0.7$, and $1.0$
%with ${v}^{\rm (n)}_{ij}={v}^{\rm (t)}_{ij}=0$,
%where
%the contact surface is treated as ``stuck'' for
%$\delta^{\rm (t)}_{ij}< \mu |\Vect{F}^{\rm (n)}_{ij}| / k^{\rm (t)}$,
%and as ``slipping'' 
%for $\delta^{\rm (t)}_{ij}\ge \mu |\Vect{F}^{\rm (n)}_{ij}| / k^{\rm (t)}$.
% The range of the ``stuck'' state is proportional to $\mu$,
% where $|\Vect{F}^{\rm (t)}_{ij}|$ increases linearly with 
% $\delta^{\rm (t)}_{ij}$,
% while $|\Vect{F}^{\rm (t)}_{ij}|$ is constant in the ``slipping'' state.
%
%\begin{figure}[htbp]
% \includegraphics[width=\linewidth]{Ft.eps}
% \caption{The tangential contact force $|\Vect{F}^{\rm (t)}_{ij}|$
% normalized with the normal contact force $|\Vect{F}^{\rm (n)}_{ij}|$
%against the tangential displacement $|\delta^{\rm (t)}_{ij}|$ normalized with
%$|\Vect{F}^{\rm (n)}_{ij}| / k^{\rm (t)}$ for $\mu = 0.1$, $0.5$, $0.7$, and $1.0$
%with ${v}^{\rm (n)}_{ij}={v}^{\rm (t)}_{ij}=0$.
%}
% \label{Ft}
%\end{figure}

In this model, we apply an oscillatory shear 
along the $y$ direction
under the Lees-Edwards boundary condition \cite{SEvans}.
The SLLOD equations of motion are used to stabilize the
uniform sheared state as
\begin{eqnarray}
\frac{d \Vect{r}_i}{dt} & = & \frac{\Vect{p}_i}{m_i} + \dot \gamma (t) r_{i,y} \hat{\Vect{x}},
\label{SLLOD:1} \\
\frac{d \Vect{p}_i}{dt} & = & \sum_{j \neq i} \Vect{F}_{ij} - 
\dot \gamma (t) p_{i,y} \hat{\Vect{x}}, \label{SLLOD:2}\\
I_i \frac{d \omega_i}{dt} & = & T_i, \label{SLLOD:3}
\end{eqnarray}
where we have introduced the time dependent 
shear rate $\dot \gamma(t)$,
the peculiar momentum $\Vect{p}_i$
defined by Eq. \eqref{SLLOD:1}, 
the unit vector parallel to the $x$-direction $\hat{\Vect{x}}$,
the mass $m_i = \pi \rho d_i^2 / 4$,
and the moment of inertia $I_i = m_i d_i^2/8$.

As an initial state, 
the disks are randomly placed in the system
with the initial packing fraction $\phi_{\rm I} = 0.75$.
After relaxing the system
to a mechanical equilibrium state,
we compress the system without shear  in small steps
until the packing fraction reaches a given value $\phi$.
In each step, 
we change the linear system size
and the position of the grain $i$ as
\begin{eqnarray}
L^{(n_s+1)} & = & L^{(n_s)} \sqrt{\frac{\phi^{(n_s)}}{\phi^{(n_s+1)}}}, \\
\Vect{r}_i^{(n_s+1)} & = & \Vect{r}_i^{(n_s)} \sqrt{\frac{\phi^{(n_s)}}{\phi^{(n_s+1)}}},
\end{eqnarray}
and relax the grains to the mechanical equilibrium state.
Here, $L^{(n_s)}$, $\Vect{r}_i^{(n_s)}$, and $\phi^{(n_s)}$
denote the system size, the position of the grains $i$, and the packing fraction
at the $n_s$-th step, respectively.
The increment of the packing fraction is defined by
$\Delta \phi \equiv \phi^{(n_s+1)} -\phi^{(n_s)}$.
Here, we regard the state for $T< T_{\rm th}$
as the mechanical equilibrium state,
where we have introduced the kinetic temperature
$T \equiv \sum_i |\Vect{p}_i|^2 / (2 m_i N)$ and a threshold $T_{\rm th}$.
It should be noted that the origin of the coordinate axes is located at the center
of the system.

Let us summarize a set of parameters
used in our simulation.
We use $k^{\rm (t)} = k^{\rm (n)}$ and $\eta^{\rm (n)} = \eta^{\rm (t)} 
= \sqrt{m_0k^{\rm (n)}}$, where $m_0$ is the mass of a grain
with the diameter $d_0$.
This set of the parameters corresponds to the constant restitution
coefficient $e=0.043$.
We adopt the leapfrog algorithm
with the time step $\Delta t = 0.05 \tau$, where $\tau$ is 
the characteristic time of the stiffness, i.e., 
$\tau = \sqrt{m_0/k^{\rm (n)}}$. 
The number $N$ of the grains
is $4000$.
We have checked that the shear modulus is
independent of $N$ for $N \ge 4000$.
We fix the parameters $T_{\rm th}=10^{-8} (k^{\rm (n)} d_0^2)$,
$\Delta \phi=10^{-4}$, and $\omega=10^{-4} \tau^{-1}$
in the main text, but we have examined the $\omega$-dependence of the complex modulus in the next section.
We have confirmed that our results in the main text
are insensitive to $T_{\rm th}, \Delta \phi$ and $\omega$
if they are sufficiently small.
%For each parameter point ($\gamma_0$, $\mu$, $\phi$),
%we average over at least $4$ independent ensembles.
Here, the storage modulus $G$ is calculated from 
Eq. (1) in the main text with the aid of the shear stress $\sigma(t)$ given by
\begin{eqnarray}
\sigma(t) &= & -\frac{1}{L^2} \sum_i^N \sum_{j>i} 
r_{ij,x}(t)
F_{ij,y}(t) \nonumber \\
& & - \frac{1}{L^2} \sum_i^N \frac{p_{i,x}(t) p_{i,y}(t)}{m_i}.
\end{eqnarray}

\section{Frequency dependence of storage and loss moduli}
\label{Omega}

\begin{figure}[htbp]
\includegraphics[width=\linewidth]{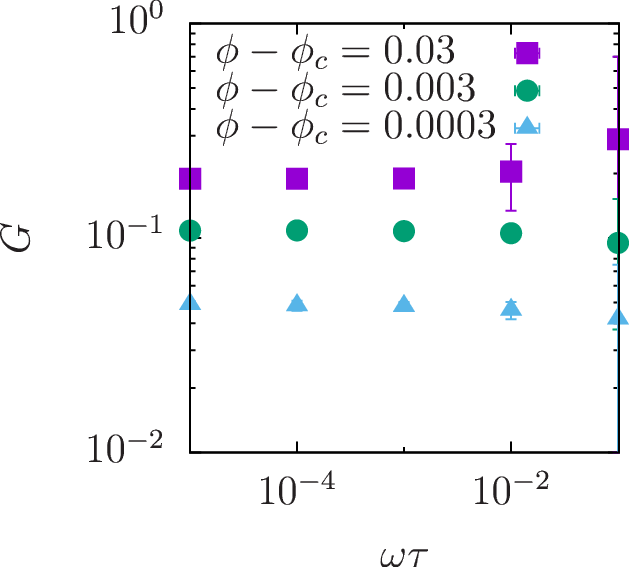}
\caption{The storage modulus $G$ against $\omega$ for $\phi-\phij=0.03, 0.003$ and $0.0003$ at $\mu=1.0$ and $\gamma_0=10^{-7}$.
}
\label{G_storage}
\end{figure}

This section deals with the dependence of the complex shear modulus
$G^*=G+i G''$ 
on $\omega$ in the linear response regime.
In Fig. \ref{G_storage}, we show 
the storage modulus $G$ against $\omega$ for various $\phi$ at $\mu=1.0$ and $\gamma_0=10^{-7}$.
As shown in Fig. \ref{G_storage},
$G$ for each $\phi$ is almost independent of $\omega$ for $\omega \tau <10^{-2}$.
Hence, we have not discussed
the $\omega$-dependence of $G$ in the main text.

\begin{figure}[htbp]
\includegraphics[width=\linewidth]{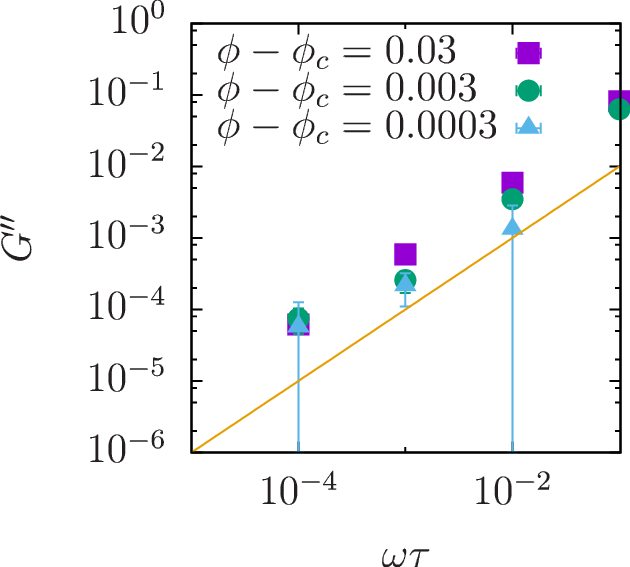}
\caption{The loss modulus $G''$ against $\omega$ for $\phi-\phij=0.03, 0.003$ and $0.0003$ at $\mu=1.0$ and $\gamma_0=10^{-7}$.
The solid line represents $G'' \sim \omega$.
}
\label{G_loss}
\end{figure}

Figure \ref{G_loss} exhibits
the loss modulus $G''$ against $\omega$ for various $\phi$ at $\mu=1.0$ and $\gamma_0=10^{-7}$.
Here, $G''$ is given by
\begin{eqnarray}
G'' =  \frac{\omega}{\pi} \int_0^{2\pi/\omega} dt
\sigma(t) \sin(\omega t)/\gamma_0.
\end{eqnarray}
As shown in Fig. \ref{G_loss},
$G''$ is almost proportional to $\omega$.
These $\omega$ dependences
of 
$G$ and $G''$ indicate that the rheological properties
in our model are essentially described by
the Kelvin-Voigt viscoelasticity.
This result is reasonable because the Cundall-Strack model 
relies on the Kelvin-Voigt model \cite{SCundall}.
Moreover, the $\phi$-dependence of $G''$ is not  clearly visible.
Hence, we have investigated only $G$ in the main text.

\section{The stress-strain curves for various $\gamma_0$}
\label{Stress}

\begin{figure}[htbp]
\includegraphics[width=\linewidth]{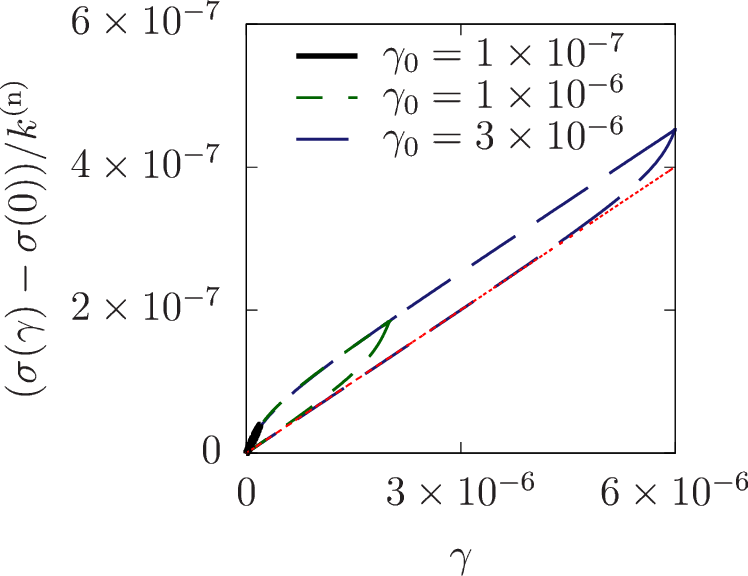}
\caption{
The intrinsic shear stress $\sigma(\gamma) - \sigma(0)$ against $\gamma$ 
for $\gamma_0 = 3.0 \times 10^{-6}$, $1.0 \times 10^{-6}$, 
and $1.0 \times 10^{-7}$
at $\mu=10^{-6}$ and $\phi = 0.87$.
The dotted line indicates $\sigma(\gamma) - \sigma(0)$ 
for $\gamma_0 =3.0 \times 10^{-6}$ at $\mu=0$ and $\phi = 0.87$.
}
\label{s_mu1.0e-06}
\end{figure}

In this section, we demonstrate the dependence of the stress-strain curve
on $\gamma_0$.
Figure \ref{s_mu1.0e-06} exhibits
the intrinsic shear stress $ \sigma(\gamma) - \sigma(0)$ against 
the shear strain $\gamma$ 
for various $\gamma_0$
at $\mu=10^{-6}$ and $\phi=0.87$.
For $\gamma_0 = 10^{-7}$, 
$\sigma(\gamma) - \sigma(0)$ is proportional to $\gamma$,
but the constant of proportionality is larger than that of frictionless grains
indicated by the dotted line.
The reason for the difference in the constant of proportionality
is that 
the increase of the tangential friction force $|\Vect{F}^{\rm (t)}_{ij}|$
in the ``stick region'' enlarges
$\sigma(\gamma) - \sigma(0) $ for $\mu>0$ 
under sufficiently small $\gamma$.
As $\gamma_0$ increases,
the stress-strain curve exhibits a nonlinear behavior:
a hysteresis loop to connect
 the first linear region for
sufficiently small $\gamma$ with 
the second linear region for $\gamma>10^{-6}$,
where the gradient of the second linear region is identical to that
of the frictionless grains.
The second linear region originates from the fact that
$|\Vect{F}^{\rm (t)}_{ij}|$ 
remains constant in the ``slip region''
and does not contribute to enlarge $\sigma(\gamma) - \sigma(0)$.

\section{Method to determine the exponents}
\label{Exp}

In this section, we explain the method to determine the exponents
in the scaling plot proposed in the main text.
Here, the exponents are determined by Levenberg-Marquardt
algorithm \cite{SNC},
where we have assumed the functional form of the scaling function as
\begin{eqnarray}
{\mathcal F}(x) & = & 
\frac{1}{1+e^{\sum_{n=0}^{N_n} A_n (\ln x)^n}}
\label{F:app}
\end{eqnarray}
with the fitting parameters $N_n=1$,
$A_0 = -1.13 \pm 0.22$, and
$A_1 = 1.13 \pm 0.02$.
We have checked the numerical estimation of $b_1$ and $b_2$
with $N_n\ge2$, but $A_n$ for $n\ge2$ are almost $0$,
implying that the exponent $c$ in Eq. (4) of the main text 
is approximately 
equal to $c \approx A_1=1.13$.

\section{Determination of transition point}
\label{Phi}

\begin{figure}[htbp]
\includegraphics[width=\linewidth]{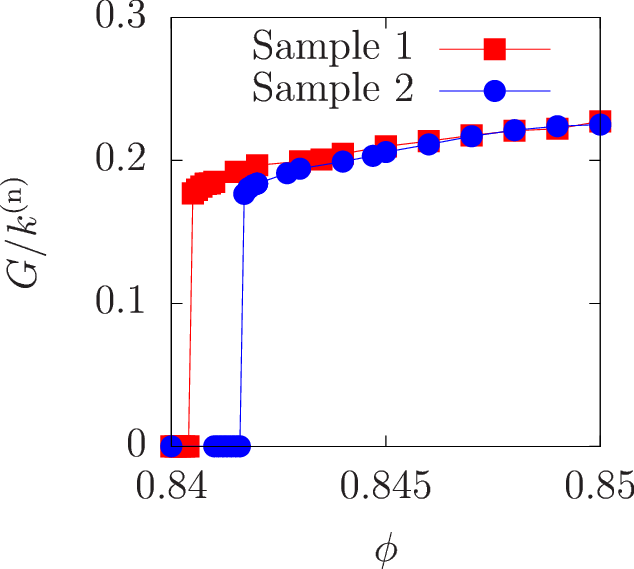}
\caption{The storage modulus $G$ against $\phi$ at $\mu=0.01$ and $\gamma_0=10^{-7}$.
``Sample 1'' and ``Sample 2'' indicate the data obtained from different
initial configurations, respectively.}
\label{G_sample}
\end{figure}

\begin{figure}[htbp]
\includegraphics[width=\linewidth]{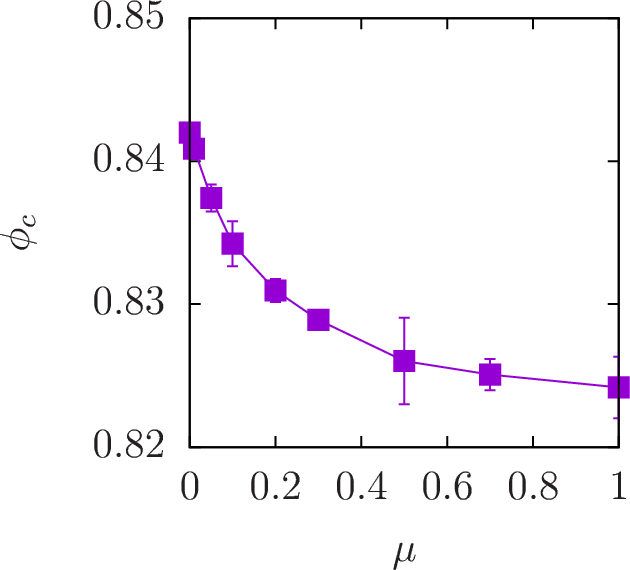}
\caption{The plots of transition points $\phij$ against$\mu$.}
\label{phij}
\end{figure}

In this section, we explain the method to determine 
the transition point $\phij(\mu)$.
To estimate $\phij(\mu)$,
we first prepare assemblies with different packing fractions
obtained from the same initial configuration of grains
with the friction coefficient $\mu$.
Then, we apply the oscillatory shear to measure 
$G$ for sufficiently small shear amplitude ($\gamma_0=10^{-7}$),
and plot $G$ against $\phi$ as shown in Fig. \ref{G_sample},
which plots
the data at $\mu=0.01$ and $\gamma_0=10^{-7}$
obtained from different initial configurations.
$G$ changes discontinuously around $\phi=0.84$,
but the critical fraction
depends on the initial configuration.
Then, we define $\phij(\mu)$ for each 
initial configuration as the minimum value of $\phi$
where $G$ for a given initial configuration exceeds 
$G_{\rm th} = 10^{-3} k^{\rm (n)}$.
In Fig. \ref{phij}, we plot the average of $\phij(\mu)$ 
against $\mu$.
$\phij(\mu)$ decreases with increasing $\mu$,
which is consistent with the previous numerical results 
\cite{SSilbert10,SOtsuki11}.

\section{The storage modulus in the linear response regime}
\label{Linear}

\begin{figure}[htbp]
\includegraphics[width=\linewidth]{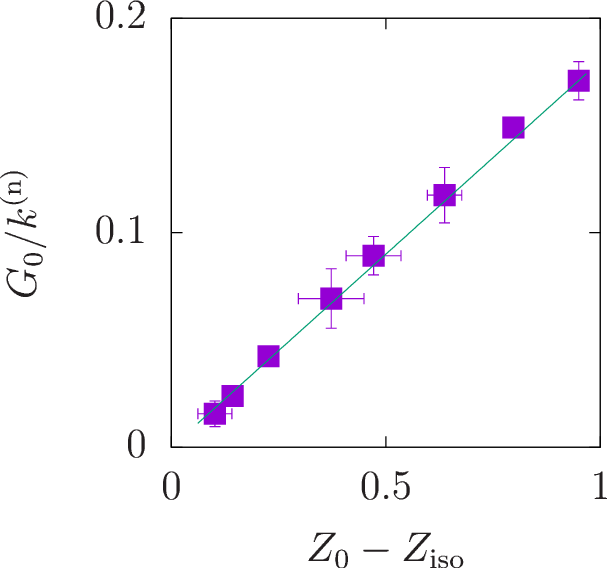}
\caption{$G_0(\mu)$
against $Z_0(\mu)-Z_{\rm iso}$ for $\mu = 0.01$, $0.05$, $0.1$, $0.2$, $0.3$, $0.5$, $0.7$, and $1.0$. 
The solid line represents Eq. \eqref{G_Z:eq}.
}
\label{G_Z}
\end{figure}

\begin{figure}[htbp]
\includegraphics[width=\linewidth]{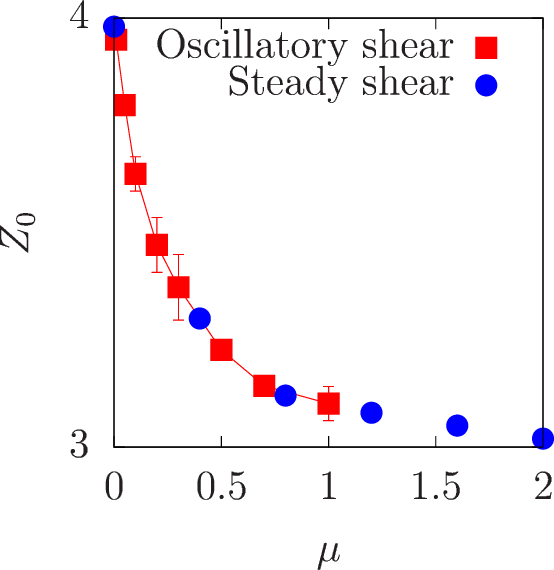}
\caption{
The coordination number $Z_0$ at the transition point $\phij$ against $\mu$
for the present system under oscillatory shear and 
the system in Ref. \cite{SOtsuki11} under steady shear.
}
\label{Z}
\end{figure}

In this section, we investigate the storage modulus 
in the linear response regime.
Figure \ref{G_Z} plots
$G_0(\mu)$
against the the excess coordination number $Z_0(\mu)-Z_{\rm iso}$ 
for various $\mu$,
where $Z_0(\mu)$ and $Z_{\rm iso}=3$ are the coordination number at $\phij$
and the isostatic value
of the coordination number for two dimensional frictional grains,
respectively.
As shown in  Ref. \cite{SSomfai},
$G_0(\mu)$ satisfies
\begin{equation}
G_0(\mu) \propto Z_0(\mu)-Z_{\rm iso}.
\label{G_Z:eq}
\end{equation}
It is known that
$Z_0(\mu)$ continuously decreases with increasing $\mu$
from the isostatic value, $4$,
for frictionless grains \cite{SSilbert10,SOtsuki11,SSomfai,SSilbert02}.
 $Z_0(\mu)$ in our model exhibits
 the identical behavior as shown in Fig. \ref{Z}.
 It should be noted that
 $Z_0(\mu)$  in our system under oscillatory shear
 is qualitatively consistent with that of
 the previous result under steady shear \cite{SOtsuki11}.
These results explain the decrease of $G_0(\mu)$
with increasing $\mu$
shown in Fig. 5 of the main text.

\begin{figure}[htbp]
\includegraphics[width=\linewidth]{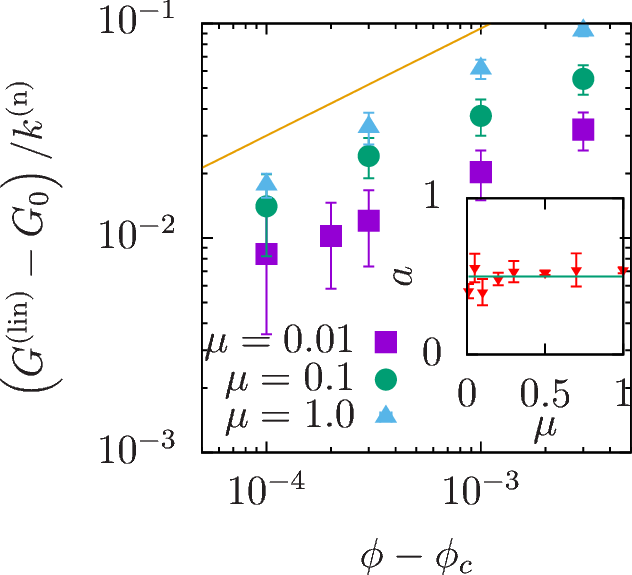}
\caption{$G^{\rm (lin)}(\mu,\phi) - G_0(\mu)$ against $\phi - \phij(\mu)$
for $\mu=0.01$, $0.05$, $0.1$, $0.5$, and $1.0$.
The solid line represents Eq. (8) in the main text with $a = 1/2$.
(Inset) The exponent $a$ evaluated by the least squares method against $\mu$.
The solid line represents $a=1/2$.
}
\label{dG_dp}
\end{figure}

We have also found that $G^{\rm (lin)}(\mu,\phi)$
satisfies Eq. (8) in the main text,
which is verified in Fig. \ref{dG_dp}.
In the inset of Fig. \ref{dG_dp}, we plot 
the exponent $a$ evaluated by the least squares method
against $\mu$,
which is almost
independent of $\mu$ and
 estimated as $a = 0.52 \pm 0.1$.

%\begin{figure}[H]


\begin{thebibliography}{99}

\bibitem{Liu} A.~J.~Liu and S.~R.~Nagel, Nature {\bf 396}, 21 (1998).

\bibitem{Hecke} M. van Hecke, J. Phys.: Condens. Matter {\bf 22}, 033101 (2009)

\bibitem{OHern02} C. S. O'Hern, S. A. Langer, A. J. Liu, and S. R. Nagel, Phys. Rev.
Lett. {\bf 88}, 075507 (2002). 

\bibitem{OHern03} C. S. O'Hern, L. E. Silbert, A. J. Liu, and S. R. Nagel, Phys. Rev. E {\bf 68}, 011306 (2003).




\bibitem{Olsson} P. Olsson and S. Teitel, Phys. Rev. Lett. \textbf{ 99}, 178001 (2007). 

\bibitem{Hatano07} T. Hatano, M. Otsuki, and S. Sasa, J. Phys. Soc. Jpn.
{\bf 76}, 023001 (2007). 

\bibitem{Hatano08} T. Hatano, 
J. Phys. Soc. Jpn. {\bf 77}, 123002 (2008).

\bibitem{Tighe} B. P. Tighe, E. Woldhuis, J. J. C. Remmers, W. van Saarloos,
and M. van Hecke,
Phys. Rev. Lett. {\bf 105}, 088303 (2010).

\bibitem{Hatano10} T. Hatano, 
Prog. Theor. Phys. Suppl. {\bf 184}, 143 (2010).


\bibitem{Otsuki08} M. Otsuki and H. Hayakawa,
Prog. Theor. Phys. {\bf 121}, 647 (2009).

\bibitem{Otsuki09} M. Otsuki and H. Hayakawa,
Phys. Rev. E {\bf 80}, 011308 (2009).

\bibitem{Otsuki10} M. Otsuki, H. Hayakawa, and S. Luding,
Prog. Theor. Phys. Suppl. {\bf 184}, 110 (2010).

\bibitem{Nordstrom} K. N. Nordstrom, E. Verneuil, P. E. Arratia, A. Basu, Z. Zhang, A. G. Yodh, J. P. Gollub, and D. J. Durian, Phys. Rev. Lett. {\bf 105}, 175701 (2010).

\bibitem{Olsson11} P. Olsson and S. Teitel,
Phys. Rev. E {\bf 83}, 030302(R) (2011).

\bibitem{Vagberg} D. V\r{a}gberg, P. Olsson, and S. Teitel,
Phys. Rev. E {\bf 83}, 031307 (2011).


\bibitem{Otsuki12}
M. Otsuki and H. Hayakawa, Prog. Theor. Phys. Suppl. {\bf 195},
129 (2012).

\bibitem{Ikeda} A. Ikeda, L. Berthier, and P. Sollich, 
Phys. Rev Lett. {\bf 109} 018301 (2012).


\bibitem{Olsson12} P. Olsson and S. Teitel, Phys. Rev. Lett. {\bf 109}, 108001 (2012).

\bibitem{DeGiuli} E. DeGiuli, G. D\"uring, E. Lerner, and M. Wyart,
Phys. Rev. E {\bf 91}, 062206 (2015).

\bibitem{Vagberg16} D. V\r{a}gberg, P. Olsson, and S. Teitel
Phys. Rev. E {\bf 93}, 052902 (2016).

\bibitem{Boyer} F. Boyer, E. Guazzelli, and O. Pouliquen,
Phys. Rev. Lett. {\bf 107}, 188301 (2011).

\bibitem{Trulsson} M. Trulsson, B. Andreotti, and P. Claudin,
Phys. Rev. Lett. {\bf 109}, 118305 (2012).

\bibitem{Andreotti} B. Andreotti, J.-L. Barrat, and C. Heussinger,
Phys. Rev. Lett. {\bf 109}, 105901 (2012).

\bibitem{Lerner} E. Lerner,  G. D\"uring, and M. Wyart,
Proc. Natl. Acad. Sci. U.S.A {\bf 109}, 4798 (2012).

\bibitem{Vagberg14}  D. V\r{a}gberg, P. Olsson, and S. Teitel
Phys. Rev. Lett. {\bf 113}, 148002 (2014).

\bibitem{Kawasaki15} T. Kawasaki, D. Coslovich, A. Ikeda, and L. Berthier,
Phys. Rev. E {\bf 91}, 012203 (2015).

\bibitem{Suzuki} K. Suzuki and H. Hayakawa,
Phys. Rev. Lett. {\bf 115}, 098001 (2015).


\bibitem{Wyart05} 
M. Wyart, Annales de Physique {\bf 30}, 3 (2005).

\bibitem{Coulais} C. Coulais, A. Seguin, and O. Dauchot,
Phys. Rev. Lett. {\bf 113}, 198001 (2014).


\bibitem{Otsuki14} M. Otsuki and H. Hayakawa, Phys. Rev. E {\bf 90},
042202 (2014).

\bibitem{Deen} M. S. van Deen, J. Simon, Z. Zeravcic, S. Dagois-Bohy,
B. P. Tighe, and M. van Hecke, Phys. Rev. E {\bf 90}, 020202(R) (2014).

\bibitem{Goodrich} C. P. Goodrich, A. J. Liu, and J. P. Sethna,
Proc. Natl. Acad. Sci. USA. {\bf 113} 9745
(2016).

\bibitem{Nakayama} D. Nakayama, H. Yoshino, and F. Zamponi,
J. Stat. Mech. 104001 (2016).

\bibitem{Boschan} J. Boschan, D. Vagberg, E. Somfai, B. P. Tighe,
Soft Matter {\bf 12}, 5450 (2016).

\bibitem{Dahmen98}
K. Dahmen, D. Erta\c{s}, and Y. Ben-Zion, Phys. Rev. E \textbf{58}, 1494 (1998).

\bibitem{Dahmen11}
K. Dahmen, Y. Ben-Zion, and J. T. Uhl, Nat. Phys. \textbf{7}, 554 (2011).






\bibitem{Otsuki11} M. Otsuki and H. Hayakawa, Phys. Rev. E 83, 051301 (2011).

\bibitem{bob_nature} D. Bi, J. Zhang, B. Chakraborty and R. Behringer, Nature {\bf 480}, 355 (2011).


\bibitem{Chialvo} S. Chialvo, J. Sun, and S. Sundaresan,
Phys. Rev. E {\bf 85}, 021305 (2012).


\bibitem{Brown} E. Brown and H. M. Jaeger, 
Phys. Rev. Lett. {\bf 103}, 086001 (2009).

\bibitem{Seto}
R. Seto, R. Mari, J. F. Morris, and M. M. Denn,
Phys. Rev. Lett. {\bf 111}, 218301 (2013).

\bibitem{Fernandez}
N. Fernandez, R. Mani, D. Rinaldi, D. Kadau, M. Mosquet, H. Lombois-Burger, J. Cayer-Barrioz, H. J. Herrmann, N. D. Spencer, and L. Isa, Phys. Rev. Lett. {\bf 111}, 108301 (2013).

\bibitem{Heussinger} C. Heussinger, Phys. Rev. E {\bf 88}, 050201 (2013).


\bibitem{Bandi} M. M. Bandi, M. K. Rivera, F. Krzakala and R.E. Ecke, Phys. Rev. E {\bf 87}, 042205 (2013).

\bibitem{Ciamarra} M. P. Ciamarra, R. Pastore, M. Nicodemi, and
A. Coniglio, Phys. Rev. E {\bf 84}, 041308 (2011).

\bibitem{Mari} R. Mari, R. Seto, J. F. Morris, and M. M. Denn,
J. Rheol. {\bf 58}, 1693 (2014).

\bibitem{Grob} M. Grob, C. Heussinger, and A. Zippelius, Phys. Rev. E
{\bf 89}, 050201 (2014).

\bibitem{Kawasaki14} T. Kawasaki, A. Ikeda, and L. Berthier, EPL {\bf 107},
28009 (2014).

\bibitem{Wyart14} M. Wyart and M. E. Cates, Phys. Rev. Lett. {\bf 112},
098302 (2014).

\bibitem{Grob16} M. Grob, A. Zippelius, and C. Heussinger, Phys. Rev. E
{\bf 93}, 030901 (2016).

\bibitem{Magnanimo} V. Magnanimo, L. La Ragione, J. T. Jenkins,
P. Wang, and H. A. Makse, EPL {\bf 81}, 34006 (2008).

\bibitem{Hayakawa} H. Hayakawa and S. Takada,
arXiv:1611.07925.







\bibitem{Somfai} E. Somfai, M. van Hecke, W. G. Ellenbroek, K. Shundyak, and W. van Saarloos, Phys. Rev. E {\bf 75}, 020301(R) (2007).


\bibitem{Cundall} P. Cundall and O. D. L. Strack, Geotechnique {\bf 29},
47 (1979).

\bibitem{Doi}
M. Doi and S. F. Edwards, \textit{The Theory of Polymer Dynamics}
(Oxford University Press, Oxford, 1990).

\bibitem{Supple} See Supplemental Material,
which includes Ref. \cite{Evans,NC,Silbert10,Silbert02}, for 
the details of the analysis and supplementary simulation results.

\bibitem{Evans}
D.~J.~Evans and G.~P.~Morriss, \textit{Statistical Mechanics of Nonequilibrium Liquids} 2nd ed.
(Cambridge University Press, Cambridge, 2008).


\bibitem{NC}
W. H. Press, S. A. Teukolsky, W. T. Vetterling, and B. P.
Flannery, \textit{Numerical Recipes}, 3rd ed. (Cambridge University
Press, Cambridge, 2007).



\bibitem{Silbert10} L. E. Silbert, Soft Matter {\bf 6}, 2918 (2010) 




\bibitem{Silbert02} L. E. Silbert, D. Erta\c{s}, G. S. Grest, T. C. Halsey,
and D. Levine, Phys. Rev. E {\bf 65}, 031304 (2002).




\bibitem{Tighe11} B. P. Tighe, Phys. Rev. Lett. {\bf 107}, 158303 (2011).

\bibitem{Lois} G. Lois, J. Blawzdziewicz, and
C. S. O'Hern, Phys. Rev. Lett. {\bf 100}, 028001 (2008).

\bibitem{Shen} T. Shen, C. S. O'Hern, M. D. Shattuck,
Phys. Rev. E {\bf 85}, 011308 (2012).

\bibitem{Kovalcinova} L. Kovalcinova, A. Goullet, and L. Kondic,
Phys. Rev. E {\bf 92}, 032204 (2015).

\bibitem{Henkes} S. Henkes, D. A. Quint, Y. Fily, and J. M. Schwarz,
Phys. Rev. Lett. {\bf 116}, 028301 (2016).

\bibitem{Torquato}
S. Torquato, \textit{Random Heterogeneous Materials: Microstructure and Macroscopic Properties}, 2nd ed. (Springer, New York, 2005).


\end{thebibliography}

\begin{thebibliography}{99}

\bibitem{SEvans}
D.~J.~Evans and G.~P.~Morriss, \textit{Statistical Mechanics of Nonequilibrium Liquids} 2nd ed.
(Cambridge University Press, Cambridge, 2008).

\bibitem{SCundall} P. Cundall and O. D. L. Strack, Geotechnique {\bf 29},
47 (1979).


\bibitem{SNC}
W. H. Press, S. A. Teukolsky, W. T. Vetterling, and B. P.
Flannery, \textit{Numerical Recipes}, 3rd ed. (Cambridge University
Press, Cambridge, 2007).



\bibitem{SSilbert10} L. E. Silbert, Soft Matter {\bf 6}, 2918 (2010) 


\bibitem{SOtsuki11} M. Otsuki and H. Hayakawa, Phys. Rev. E 83, 051301 (2011).


\bibitem{SSomfai} E. Somfai, M. van Hecke, W. G. Ellenbroek, K. Shundyak, and W. van Saarloos, Phys. Rev. E {\bf 75}, 020301(R) (2007).


\bibitem{SSilbert02} L. E. Silbert, D. Erta\c{s}, G. S. Grest, T. C. Halsey,
and D. Levine, Phys. Rev. E {\bf 65}, 031304 (2002).



\end{thebibliography}
\end{document}